\documentclass{elsart}
\usepackage{graphicx,epsfig}

\newcommand{\ih}{\textit{h}}
\newcommand{\ip}{\textit{p}}
\newcommand{\ic}{\textit{c}}
\newcommand{\ib}{\textit{b}}

\newcommand{\bc}{\textit{b-c}}
\newcommand{\bh}{\textit{b-h}}
\newcommand{\bp}{\textit{b-p}}
\newcommand{\ch}{\textit{c-h}}
\newcommand{\cp}{\textit{c-p}}
\newcommand{\hp}{\textit{h-p}}

\begin{document}
\begin{frontmatter}

\title{DFT study of Rb/Si(100)-2$\times$1 System}

\author[balikesir]{E. Mete\corauthref{cor}},
\corauth[cor]{Corresponding author.}
\ead{emete@balikesir.edu.tr}
\author[metu]{R. Shaltaf} and
\author[metu]{\c{S}. Ellialt{\i}o\u{g}lu}

\address[balikesir]{Physics Department, University of Bal{\i}kesir,
10100 Bal{\i}kesir, Turkey}
\address[metu]{Department of Physics, Middle East Technical University, Ankara
06531, Turkey}

\begin{abstract}
We have investigated Rb adsorption on the Si(100) surface for 0.5 and 1
monolayer coverages using the total energy method with norm-conserving
pseudopotentials.  For 2$\times$1 reconstruction at 1 ML coverage symmetrized
dimers are found to be energetically more favorable. On the other hand, half a
ML coverage is found to have symmetrical dimers only for the most stable
adsorption model. All possible surface-adatom configurations have been considered
in the calculations to find which adsorption sites are energetically favored.
In addition to the structural properties, the interface is investigated
electronically for the work function and surface states. The results are
discussed and compared with the existing experimental findings.
\end{abstract}

\end{frontmatter}

\section{Introduction}

Silicon surfaces covered with alkali metal (AM) atoms have been investigated
extensively for various coverage models. The motivation behind these studies
is to understand the metal-semiconductor junctions and to study metallization
of semiconductor surfaces. One may expect alkali metal adsorbed silicon
surfaces to exhibit similar characteristics. However, systematic differences
arise as one goes from lighter atoms to heavier ones leading to different
electronic behavior. The discussion, especially, on the atomic and electronic
structure of alkali metal adsorbed Si(100)-2$\times$1 surface is still under
debate which concerns the matters such as the adsorption sites, the saturation
coverage at room temperature (RT), the nature of alkali metal-Si bond and the
charge transfer from the adsorbate to the substrate.

Another reason why AM adsorption on Si surfaces attracted much interest is that
AM's have simple electronic structures and do not intermix with the
surface Si atoms, except Li, allowing us to investigate a relatively simpler
interface. Therefore Si-AM systems were seen as a prototype to understand the
physics of AM overlayers on semiconductor surfaces. In the last years the
advances in the engineering of nano scale structures allow device
designs based upon these contacts on the atomic scale.

\begin{figure*}[ht]
\begin{center}
\epsfig{file=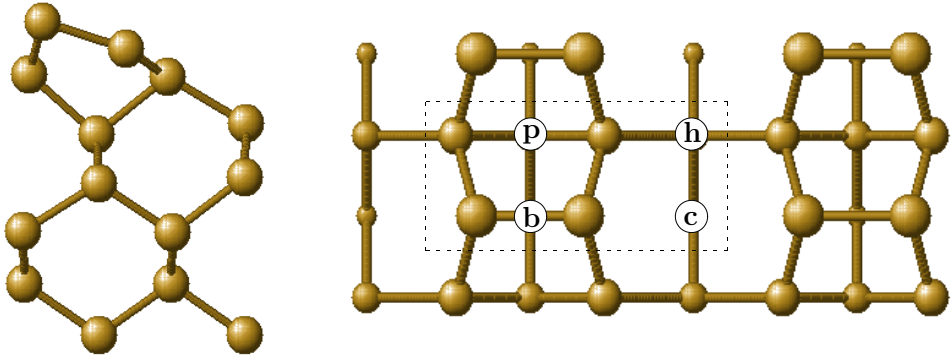,clip=true}
\caption{On the left : Schematics are shown for the side view of relaxed clean
Si(100)-2$\times$1 surface. On the right : Top view of Si(100) surface
with all possible stable adsorption sites are illustrated.
(Dimers are shown as symmetrical for visual convenience.)\label{figure1}}
\end{center}
\end{figure*}

There are, fundamentally, two proposed coverage models which have been
experimentally realized. These are 0.5 and 1 ML coverages. Half a ML model
has been first studied by Levine for Cs/Si(100)-2$\times$1 system~\cite{levine}.
In this model alkali metal atoms are placed in the middle of neighboring dimers
in the same row. This adsorption location is called as pedestal site, \ip,
(see Figure~\ref{figure1}). Alkali metal atoms adsorbed on pedestal site form a
linear chain along the dimer row preserving 2$\times$1 surface symmetry.

In recent years, there is much interest towards larger alkali metals
(Rb and Cs)~\cite{johansson,chao2,sherman,kim}. Etel\"{a}niemi et al. studied the
adsorption sites of Rb adsorbates on Si(100) surface with XSW experiments for various
coverages~\cite{etelaniemi}. They attempted to define the surface geometry at
the saturation coverage. Castrucci et al. determined AM-Si bond length and the Rb site
distribution on Si(001)2$\times$1 surface for low (0.19 $\pm$ 0.02) coverages using
XSW~\cite{castrucci}. Chao et al. investigated the surface electronic structure and
work function shifts as a function of the Rb coverage.~\cite{chao1,chao2}.
Johansson et al. made a systematic study of the electronic structure for Rb adsorption
on Si(100)2$\times$1 surface with ARUPS and IPES by comparing to the results of other
AM's~\cite{johansson}.

In this work our aim is to study Rb adsorbed Si(100) surface with 2$\times$1
periodicity for 0.5 and 1 ML coverages which include experimentally realized
adsorption sites. Main focuses of this study are to determine the atomic structure of Si
surface reconstructed with Rb coverage, to discuss the saturation coverage with
an emphasis on the Si dangling bonds and surface work function, and to investigate
the electronic structure to identify surface states. This is, to the best of our
knowledge, the first theoretical work that investigates the Rb adsorption on
Si(100) surface.

\section{Computational Details}

The calculations were performed using the plane-wave expansion approach within
the local density approximation to density functional theory as implemented in
the ABINIT code~\cite{abinit}. We generated Troullier-Martins
pseudopotentials~\cite{TM} using fhi98pp~\cite{fhi98pp} package with
PW92~\cite{pw92} exchange-correlation model. We treated the true electronic
valence states in the calculations. Nonlinear core-valence corrections were
taken into account for the case of Rb by setting the cut-off radius
to 3.6 bohr. To test Rb pseudopotential, Rb crystals having bcc structure have
been reproduced with the lattice parameter 5.49 {\AA}. This value is 1.6 \%
lower than the experimental value of 5.58 \AA~\cite{barrett}.

We carefully tested the convergence behavior of our calculations with respect both
to the plane-wave cut-off and to the k-point grid. A series of tests showed that
8$\times$4$\times$1 Monkhorst-Pack mesh~\cite{monkhorst-pack} with 16 Ryd cut-off
gives well converged results for our self-consistent calculations. In other words,
any further increase in these parameters would not enhance the accuracy of the
calculated values like lattice constants, band structures etc. The convergence
criterion to reach the lowest energy configuration is determined by minimizing the
forces using the Hellmann-Feynman theorem.

For surface geometry and band structure calculations we have used a slab with
8 layers of Si. In order to prevent an interaction between the two surfaces, the
bulk dangling bonds are hydrogenated and a vacuum region of 10 {\AA} is used .
In geometry optimization calculations, Si atoms which stay at the two bottom layers
are kept frozen and all the other atoms are let relax to their minimum energy
configurations. In order to observe if the surface symmetry breaks we did not
impose any symmetry operations other than the identity which naturally incorporates.

We performed separate self-consistent calculations to determine surface work
function using a symmetrical slab having a thickness equal to 10 Si layers and being
separated from each other with a vacuum region of $\sim$ 10 \AA. After relaxing the
clean surface into minimum energy geometry we freezed 4 Si atoms which are located
in the middle of this slab. We also repeated geometry optimizations using symmetrical
slab for all of the coverage models as done in the case of hydrogenated surface. These
calculations produced equivalent results.

\section{Results and Discussion}

Although a number of experimental studies are present for Rb adsorption on
Si(100) surface, no first principles work has been published yet. This paper
not only studies Rb adsorption on Si(100) theoretically, but also presents calculated
results like Rb-Si bond length, adsorption energy, etc. which have not been measured
experimentally. In a few papers these values are estimated from the theoretical
studies for K/Si(001) or Cs/Si(001) systems. We tabulated our calculated results
for structural properties in Table~\ref{table1}. For the surface energetics we
used the formula as described in the reference~\cite{shaltaf} to calculate the
adsorption energies. The adsorption energy represents the negative of the binding
energy.

\vskip2mm
\begin{table*}[htb]
\begin{center}
\caption{Calculated parameters for the atomic structure for 0.5 ML and
1 ML coverages of Rb adsorption on Si(100)2$\times$1 system. The lengths,
$d_{\rm dimer}$, $d_{\rm Rb-dimer}$ and $\Delta x$, are all in \AA.
$d_{\rm Rb-dimer}$ is the distance from the adsorbed Rb atom to the
dimer center. $\Delta x$ is the height of the Rb atom measeured from the
dimer center.\label{table1}}
\vskip2mm
\begin{tabular}{ccccccc}\hline
$\Theta$ & model & $d_{\rm dimer}$ & $d_{\rm Rb-dimer}$ & $\Delta x$ &
$\alpha_{\rm tilting}$ & $E_{ad}$ (eV) \\ \hline
0.5 & \ib & 2.37 & 3.05 & 3.01 & 10.4 & 0.74 \\
& \ic & 2.30 & 4.12 & 1.67 & 5.6 & 1.21 \\
& \ip & 2.39 & 3.04 & 2.36 & 8.0 & 1.28 \\
& \ih & 2.32 & 4.42 & 1.26 & 2.4 & 1.69 \\ \hline
1 & \bp & 2.44 & 2.75-3.02 & 2.75-2.34 & - & 0.50 \\
& \ch & 2.34 & 4.03-4.45 & 1.30-1.27 & - & 0.68 \\
& \bc & 2.40 & 3.03-4.16 & 3.04-1.67 & - & 1.11 \\
& \cp & 2.42 & 4.20-3.01 & 1.76-2.34 & - & 1.39 \\
& \bh & 2.48 & 2.98-4.42 & 2.98-1.15 & - & 1.46 \\
& \hp & 2.49 & 4.41-3.00 & 1.11-2.32 & - & 1.72 \\ \hline
\end{tabular}
\end{center}
\end{table*}
\vskip2mm

Our calculated clean surface dimer length is 2.27 \AA. It compares well with
the experimental value of 2.20 $\pm$ 0.05~\cite{mangat}. This value becomes
larger because of the adsorption which causes Rb to make a charge transfer to
the closest dimer Si atom. In turn, this saturation of the dangling bonds on the
surface leads to a stretching of the dimer. When we compare the dimer lengths
between 0.5 ML and 1 ML coverages, we see that the dimer is more stretched for
1 ML adsorption models.


For 0.5 ML coverage, bridge (\ib), cave (\ic), hollow (\ih) and pedestal (\ip) sites
are found to be stable adsorption configurations. We tested the stability of two
more sites which are the top of one of the dimer Si atoms and the site on the half
way from \ip~to \ih, so called to shallow site. None of these two sites were found
to be stable. Each time we put an Rb atom on these locations they relaxed into the
nearest stable sites. Energetically the most favorable adsorption site is \ih. The
total energy of this geometry is 0.41 eV lower than the second lowest configuration
\ip. This surface geometry has the dimer length of 2.32 \AA~with a tilting angle
of 2.4$^\circ$. Symmetrical dimers which correspond to a tilting angle of less
than 5$^\circ$ are found only for this model. The other adsorption configurations -
\ib, \ic~and \ip - result in asymmetrical dimers.

\vskip2mm
\begin{figure*}[ht]
\begin{center}
\epsfig{file=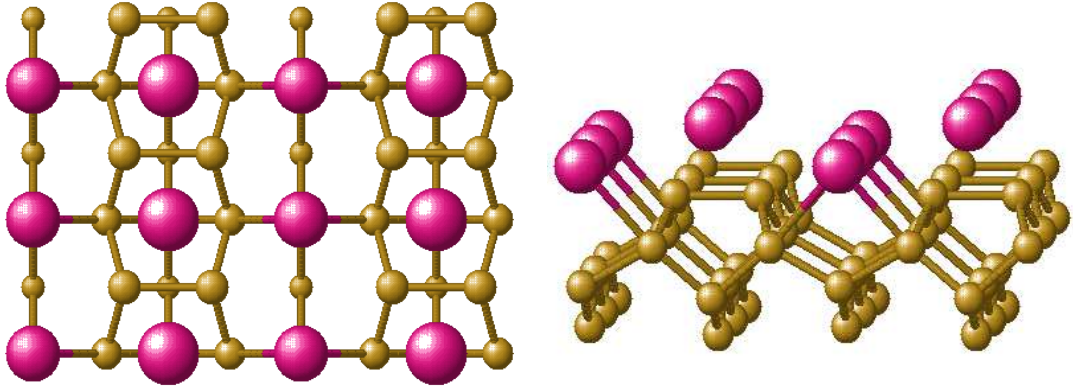,clip=true}
\caption{On the left : Illustration for top view of \hp~(or double layer) adsorption
model on Si(100)-2$\times$1 surface. (Upper is shown bigger.) On the right : Side
view of Rb/Si(100)-2$\times$1 system for 1 ML \hp~configuration.\label{figure2}}
\end{center}
\end{figure*}

For 1 ML coverage, we have studied all the possible adsorption configurations and have
found bridge-cave (\bc), bridge-hollow (\bh), bridge-pedestal (\bp), cave-pedestal (\cp),
cave-hollow (\ch) and hollow-pedestal (\hp) are the stable site combinations. Any
combination with the top and/or shallow sites which are as described in the case of
half ML, are energetically unstable. Based on an estimation for the Rb-Si bond distance
Etel{\"a}niemi et.al.\cite{etelaniemi} argue that two nearest neighbor adsorption
sites cannot be occupied at the same time. We found that these configurations
(\bp~and \ch) produce the lowest two adsorption energies. Therefore, they are
energetically the least probable adsorption models for 1 ML coverage.

The 2$\times$1 symmetry was found to be preserved
for these 1 ML adsorption coverage models as suggested by the LEED result of
Chao et al.~\cite{chao1,chao2} and ARUPS result of Johansson et al.~\cite{johansson}.
Among these \hp~is the energetically most favorable configuration which corresponds
to PV model of Etel{\"a}niemi et.al.\cite{etelaniemi}. In our calculations \hp~has a
total energy of 0.52 eV less than the second lowest energy coverage model \bh. Rb-Si bond
length for this configuration has been found to be 3.20 \AA~and 3.24 \AA~measured from Rb
atoms at \ih-site and \ip-site, respectively, to the nearest Si atom. In this geometry
symmetrically dimerized surface is energetically favored. After the relaxation the
dimer length is measured to be 2.49 \AA~which is close to the experimental value of
2.3 - 2.4 \AA~reported by Etel{\"a}niemi et al.\cite{etelaniemi}. This adsorption
model is also referred as the double layer model. Rb atoms adsorbed at \ih~and \ip-sites
are separated a distance 1.21~\AA~along [100] as shown in Figure~\ref{figure2}. This
difference in the heights is close to one Si layer which has the value $\sim$1.20 - 1.30~\AA.
In this model the nearest-neighbor distances for the Rb atoms have been predicted by
Johansson et al. to be 3.84 - 4.0 \AA~\cite{johansson} which is in a very good agreement
with our calculated values of 3.82~\AA~along [011] and 4.0~\AA~along [0$\bar 1$1].

The work function is calculated as the difference between the vacuum level and the
Fermi energy ($E_{\rm F}$). The vacuum level is determined from the self-consistent,
plane-averaged potential ($V_{\rm av}$) in the middle of the symmetrical slabs along
the direction [100] perpendicular to the surfaces. It can be obtained from the Poisson's
equation,
\begin{equation}
{\partial^2\over\partial x^2}V_{\rm av}(x)=-4\pi\rho_{\rm av}(x)
\end{equation}
and is given by the relation,
\begin{equation}
V_{\rm av}(x) = -4\pi\int_x^\infty \rho_{\rm av}(x')x'dx'+4\pi x\int_x^\infty \rho_{\rm av}(x')dx'
\end{equation}
where
\begin{equation}
\rho_{\rm av}(x)={1\over A}\int_{\,\,\,A}\!\!\int \vert\psi(x,y,z)\vert^2 dydz\,.
\end{equation}

\begin{figure*}[ht]
\begin{center}
\epsfig{file=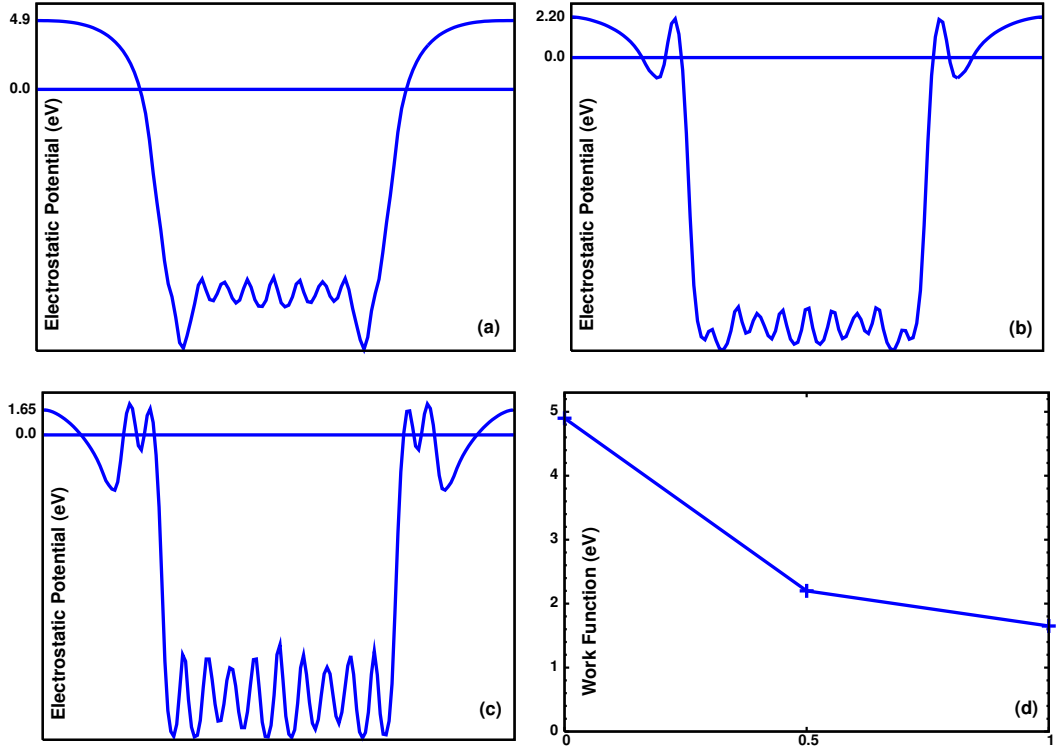,width=14cm,clip=true}
\end{center}
\caption{Planar-averaged electrostatic potential $V(x)$ (a) for the clean relaxed geometry,
(b) for $h$ adsorption model and (c) for \hp~ model along [100], the direction perpendicular
to the surface within the cell. (d) Calculated shifts in work function as a function of the
coverage.\label{figure3}}
\end{figure*}

Calculated values for the plane-averaged electrostatic potential and the work function
corresponding to different adsorption models are shown in the Figure~\ref{figure3} by
setting the zero level at the Fermi energy. Figure~\ref{figure3}(a) represents clean surface.
The work function for Si(100)-2$\times$1 surface was reported as 4.9 eV by Abukawa et al.\cite{abukawa}.
It agrees very well to our theoretical result of 4.9 eV, for the clean relaxed surface.
Figure~\ref{figure3}(b) corresponds to the half ML model of Levine~\cite{levine} which is
the $h$-model in our case. For 1 ML coverage we presented the result in Figure~\ref{figure3}(c)
which represents PV model of Etel{\"a}niemi et.al.\cite{etelaniemi} which is the \hp~model in
our calculations. The zigzag regions in the middle of these figures represent Si layers in the
bulk. A Si atom layer sits in each valley and a localized charge coincides with each peak in
this bulk region. Each peak above the Fermi level represents the position of a Rb atom. The
tales of the electrostatic potential away from the bulk region go into the vacuum regions.

\vskip2mm

The work functions were calculated as 2.20 eV and 1.65 eV for the cases of 0.5 ML and 1 ML,
respectively. When compared with the results of Johansson et al.~\cite{johansson} and
Chao et al.~\cite{chao2}, we obtained a similar behavior for the change in the work
function as Rb coverage increases. These work function shifts as a function
of coverage are shown in the Figure~\ref{figure3}(d). The lowering in the work function
is more rapid between the clean surface and 0.5 ML coverage than that of between 0.5 ML and
1 ML coverages. This behavior is consistent with the experimental observations~\cite{johansson,chao2}.
Moreover, they reported that the work function takes a constant value after the saturation
coverage. This can be explained by the fact that there is no further lowering in the work
function by Rb adsorption after the saturation reached for Si(100) surface. It appears from the
Figure~\ref{figure3}(d) that 0.5 ML cannot be the saturation coverage for Rb/Si(100)-2$\times$1
system since there is a shift in the work function for the coverages greater than half a ML.
We calculated the change in the work function for 1 ML to be 3.25 eV. The experimental result
for the saturation coverage is 3.4 eV given by Johansson et al.~\cite{johansson} and
Chao et al.~\cite{chao2}. The comparison of our theoretical results with the experimental
findings gives a strong evidence that the saturation coverage occurs at about 1 ML.

\begin{figure*}[ht]
\begin{center}
\epsfig{file=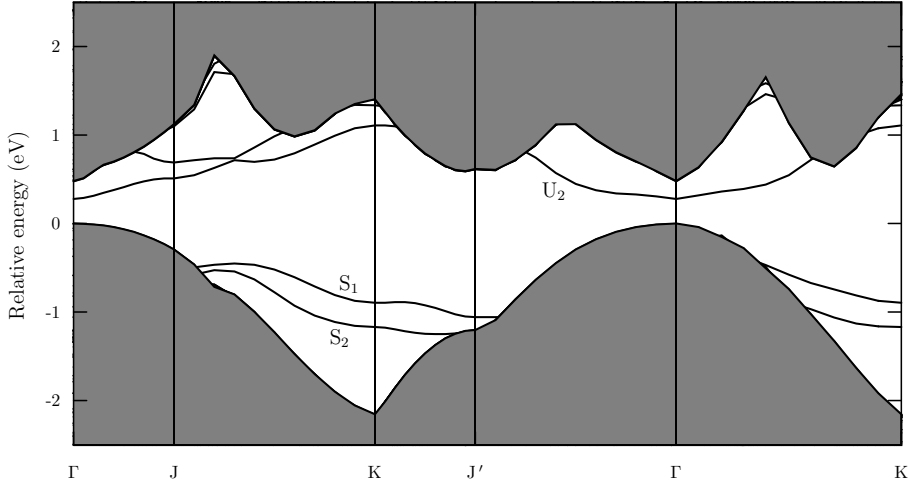,width=12cm,clip=true}
\end{center}
\caption{Bulk projected surface energy bands. The shaded region is the
projected bulk band continuum.\label{figure4}}
\end{figure*}

The bulk projected surface band structure is presented in Figure~\ref{figure4}. Energy
values are given relative to the valence band maximum (VBM) along high symmetry points.
The direct band gap is calculated to be 0.47 eV which is underestimated by LDA as expected.
The experimental value was given as 0.6 eV by Johansson et al.~\cite{johansson}.
Conduction band minimum (CBM) and valence band maximum (VBM) both occur at $\Gamma$.
The overall electronic behavior reveals similar characteristics to the
Si(100)2$\times$1-AM band structure except the dispersion and location of surface states.
We identified surface states which fall into the band gap. Two unoccupied and two occupied
surface states are shown in Figure~\ref{figure4}.

\vskip2mm

Chao et al., in their ARUPS study, reported two Rb-induced surface states $C1$ and $C2$ for
a RT saturated Si(100)-2$\times$1 surface with energies of -0.55 and -1.55 eV at $\Gamma$,
respectively~\cite{chao2}. Their results are in favor of the proposed double-layer model
which leads to a semiconducting surface. The states $C1$ and $C2$ correspond to the occupied
states S$_1$ and S$_2$ in our case with the energies -0.35 and -0.72 eV at $\Gamma$.

Johansson et al. observed a Rb-overlayer derived empty band, $U_2$, with a large isotropic
dispersion~\cite{johansson}. They argued that even though they observed a metallization due to
the large dispersion of the unoccupied band $U_2$ at saturation coverage, the surface electronic
behavior is expected to be semiconducting at 1 ML with a surface band gap of about 0.6 eV as
they measured. The explanation of the occupation of this empty band was given by the presence of
small amounts of excess charge on the surface together with the close energy position of
$U_2$ at VBM. The minimum of this band was reported to be about 0.2 eV above the VBM at
$\Gamma$~\cite{johansson}. There is a good agreement with this experimental value such
that we found the separation between the $U_2$ minimum and the VBM to be 0.25 eV. Although
they observed a metallization at saturation coverage we see from the Figure~\ref{figure4} that
Rb/Si(100)-2$\times$1 system has a semiconducting electronic structure at 1 ML for the \hp-model
even we see a large dispersion of 0.84 eV in the conduction band $U_2$.

\section{Conclusion}

In the case of 0.5 and 1 ML coverages, all possible configurations were
studied for Rb adsorption on Si(100) surface having p(2x1) symmetry.
For 0.5 ML coverage, \ih-site and for 1ML coverage, \hp~model are found to be the
minimum energy configurations. Our results suggest that Rb atoms prefer to be
adsorbed firstly, on \ih-site then on \ip-site as the coverage increases.

Clean surface work function was calculated to be 4.9 eV which is equivalent to
the experimental value of Abukawa et al~\cite{abukawa}. The work functions for
0.5 ML and 1 ML are 2.20 eV and 1.65 eV, respectively. A 3.25 eV shift in the
work function was found for \hp~model which agrees well with the experimentally
observed shift of 3.4 eV for the saturation coverage at RT.
Bulk-projected surface bands were calculated within the affinity of the
gap for the case of \hp~model. Experimentally resolved $S_1$ and $S_2$ bands
together with the unoccupied $U_2$ band are identified as surface states.
1 ML Rb adsorbed Si(100)-2$\times$1 surface was electronically found to be
semiconducting. Being consistent with the experiments, the separation at
$\Gamma$ was found to be 0.25 eV between VBM and Rb driven band $U_2$. When
the change in work function and the energetics of the surfaces are compared,
our results support the double-layer model (\hp) as the saturation coverage.
Our theoretical results fit well to the picture drawn by the experimental findings.

\section{Acknowledgement}

This work was supported by T{\"U}B\.{I}TAK, The Scientific and Technical
Research Council of Turkey, Grant No. TBAG-2036 (101T058).

\end{document}